\documentclass[a4paper,11pt]{article}
\usepackage{pos}

\usepackage{subfigure}
\DeclareMathOperator{\Tr}{Tr}

\title{Thermal phase transitions in rotating QCD with dynamical quarks}
\ShortTitle{Thermal phase transitions in rotating QCD with dynamical quarks}

\author[a]{V.~V.~Braguta}
\author[a]{A.~Yu.~Kotov}
\author*[a]{A.~A.~Roenko}
\author[a,b]{D.~A.~Sychev}

\affiliation[a]{ Bogoliubov Laboratory of Theoretical Physics, Joint Institute for Nuclear Research, Dubna, 141980 Russia }
\affiliation[b]{ Moscow Institute of Physics and Technology, Dolgoprudny, 141700 Russia }

\emailAdd{vvbraguta@theor.jinr.ru}
\emailAdd{akotov@theor.jinr.ru}
\emailAdd{roenko@theor.jinr.ru}
\emailAdd{sychev.da@phystech.edu}

\abstract{Relativistic rotation causes a change of QCD critical temperatures. Various phenomenological and effective models predict a decrease of the critical temperatures in rotating QCD. Nevertheless, lattice simulations showed that the critical temperature in gluodynamics increases due to rotation. We extend the lattice study to the theory with dynamical fermions.
We present the first lattice results for rotating QCD with  $N_f=2$ dynamical clover-improved Wilson quarks.
We also study separately the effect of rotation on gluonic and fermionic degrees of freedom. 
It is shown that separate rotations of gluons and fermions have opposite effects on the critical temperatures. In aggregate, the pseudo-critical temperatures in QCD increase with angular velocity. Dependence of the results on the pion mass is also discussed.
}

\FullConference{
The 39th International Symposium on Lattice Field Theory,\\
8th-13th August, 2022,\\
Rheinische Friedrich-Wilhelms-Universität Bonn, Bonn, Germany
}

\begin{document}
\maketitle

\section{Introduction}
In non-central heavy-ion collisions a created droplet of quark-gluon plasma (QGP) is expected to have non-zero angular momentum ~\cite{Jiang:2016woz, Becattini:2007sr, Baznat:2013zx, STAR:2017ckg}. Hydrodynamic simulations predict the vorticity of rotating QGP to reach values of 20-40~MeV (0.1-0.2 fm$^{-1}$)~\cite{Jiang:2016woz}.
From the experimental data on $\Lambda,\bar\Lambda$-hyperon polarization the average vorticity of created quark-gluon matter may be estimated as 
6~MeV~\cite{STAR:2017ckg}. Relativistic rotation of QGP causes various interesting phenomena, well-known examples of which include the chiral vortical effect~\cite{Kharzeev:2015znc, Vilenkin:1979ui, Prokhorov:2018bql, Prokhorov:2018qhq} and a polarization of created particles~\cite{Rogachevsky:2010ys, Teryaev:2017wlm}.
The critical temperature in QCD is also affected by relativistic rotation.

The behavior of QCD critical temperatures in the case of rotation was actively investigated via different approaches (see \cite{Jiang:2016wvv, Chernodub:2016kxh, Wang:2018sur, Ebihara:2016fwa, Golubtsova:2021agl, Golubtsova:2022ldm, Chen:2020ath, Chernodub:2020qah,Fujimoto:2021xix, Chernodub:2022qlz, Chen:2022smf, Chernodub:2022wsw, Chernodub:2022veq} and refs. therein), and most of them predict a decrease of the critical temperatures. It should be noted that relativistic rotation affects both fermions and gluons in QCD, and an interplay between strong interaction and rotational motion may lead to non-trivial results. The chiral symmetry breaking critical temperature in the Nambu-Jona-Lasinio model (NJL), which is mostly focused on fermionic degrees of freedom, decreases with angular velocity due to suppression of the chiral condensate~\cite{Jiang:2016wvv}. Our results for rotating gluodynamics are opposite~\cite{Braguta:2020eis, Braguta:2021jgn, Braguta:2021ucr}. Recently it was shown that in the 
NJL model the contribution of rotating gluons can be taken into account via running of the effective coupling, which changes the behavior of the critical temperature in the rotating system~\cite{Jiang:2021izj}.

In this paper we present our first results for critical temperatures in rotating QCD with $N_f = 2$ dynamical clover-improved Wilson fermions.
We demonstrate that the critical temperatures tend to increase due to rotation and discuss the dependence of the results on the pion mass. We also consider several regimes of rotation and show that separate rotations of gluons and fermions influence the critical temperatures in an opposite way.

\section{Rotating reference frame}
In order to carry out the lattice study of the rotating system we use the approach developed in Refs.~\cite{Yamamoto:2013zwa, Braguta:2020eis, Braguta:2021jgn, Braguta:2021ucr}. Namely, QCD at thermal equilibrium is investigated in the reference frame which rotates with the system.
In this reference frame the rotation can be represented in terms of an external (static) gravitational field.
Using the standard technique one can build the partition function, which is determined through the action of the system in Euclidean space~\cite{Yamamoto:2013zwa}. Unfortunately, due to the sign problem~\cite{Yamamoto:2013zwa, Braguta:2020eis, Braguta:2021jgn, Braguta:2021ucr} direct Monte-Carlo simulation of the rotating system is not possible today. To overcome this problem we conduct simulations with imaginary angular velocity $\Omega_I = -i\Omega$ and analytically continue the results to real angular velocity.
In rotating coordinates the Euclidean metric tensor takes the form
\begin{equation}\label{eq:metricE}
g^E_{\mu \nu} = 
\begin{pmatrix}
1 & 0 & 0 & y \Omega_I \\
0 & 1 & 0 & -x\Omega_I  \\ 
0 & 0 & 1 & 0 \\
y\Omega_I  & -x\Omega_I & 0 & 1 + r^2 \Omega_I^2
\end{pmatrix}\, ,
\end{equation}
where $r=\sqrt{x^2+y^2}$ denotes distance from the rotational axis (assumed to be $z$-axis). 

In general, continuum action for gluons and quarks in an external gravitational field may be written as
\begin{gather}\label{eq:Sg_action}
S_{G} = \frac{1}{4 g^{2}} \int\! d^{4} x\, \sqrt{g_E}\,  g_E^{\mu \nu} g_E^{\alpha \beta} F_{\mu \alpha}^{a} F_{\nu \beta}^{a} \, , \\
\label{eq:Sf_action}
S_{F} = \int\! d^{4} x\, \sqrt{g_E}\,  \bar\psi \big( \gamma^\mu (D_\mu - \Gamma_\mu) + m \big)\psi \, ,
\end{gather}
where $D_\mu = \partial_\mu - i A_\mu$ is the covariant derivative,  $\Gamma_\mu$ is the spinor affine connection, which is defined in terms of a vierbien. 
To construct the Dirac operator with the metric tensor~\eqref{eq:metricE} we use the vierbein in the form as in Ref.~\cite{Yamamoto:2013zwa}.

\section{Lattice setup}
In present study we discretize non-rotating parts of action using RG-improved (Iwasaki) lattice gauge action~\cite{Iwasaki:1985we} and $N_f=2$ clover-improved Wilson fermions~\cite{Sheikholeslami:1985ij}.
Relativistic rotation is introduced according to Ref.~\cite{Yamamoto:2013zwa}; 
thus, the lattice gauge action has the following form:
\begin{multline}\label{eq:rot_action_lat_imp_G}
S_{G} = \beta  \sum_{x}\Big( (c_0 + r^{2}\Omega_I^{2}) W^{1\times1}_{xy} + (c_0 + y^{2}\Omega_I^{2}) W^{1\times1}_{xz}+ (c_0 + x^{2}\Omega_I^{2}) W^{1\times1}_{yz} +
c_0 \big(W^{1\times1}_{x\tau} + W^{1\times1}_{y\tau} + W^{1\times1}_{z\tau}\big) +{} \\
{} +   y\Omega_I \big(W^{1\times1\times1}_{x y \tau} + W^{1\times1\times1}_{x z \tau}\big)
- x\Omega_I \big( W^{1\times1\times1}_{y x \tau} +  W^{1\times1\times1}_{y z \tau}\big) 
+ x y\Omega_I^{2}  W^{1\times1\times1}_{x z y} + \sum_{\mu\neq\nu} c_1 W^{1\times2}_{\mu\nu} \Big) \, ,
\end{multline}
with $\beta = 6/g^2$,  $c_0 = 1 - 8 c_1$ and $c_1 = -0.331$, where
\begin{gather}
    W^{1\times1}_{\mu\nu}(x) = 1 - \frac{1}{3} \text{Re} \Tr \ \bar{U}_{\mu\nu}(x)\,, \\
    W^{1\times2}_{\mu\nu}(x) = 1 - \frac{1}{3} \text{Re} \Tr \ R_{\mu\nu}(x)\,, \\
    W^{1\times1\times1}_{\mu\nu\rho}(x) = - \frac{1}{3} \text{Re} \Tr \ \bar{V}_{\mu\nu\rho}(x)\,,
\end{gather}
and $\bar{U}_{\mu\nu}(x)$ denotes the clover-type average of four plaquettes, 
$R_{\mu\nu}(x)$ is the rectangular loop, 
$\bar{V}_{\mu\nu\rho}(x)$ is the asymmetric chair-type average of eight chairs~\cite{Braguta:2021jgn}.
The quark lattice action is
\begin{equation}\label{eq:action_F}
    S_F = \sum_{f=u,d}\sum_{x_1, x_2} \bar\psi^{f}(x_1) M_{x_1,x_2} \psi^{f}(x_2)\,,
\end{equation}
\vspace{-1.5em}
\begin{multline*}
    M_{x_1, x_2} = \delta_{x_1,x_2} - \kappa \bigg[(1-\gamma^x) T_{x+} + (1+\gamma^x) T_{x-} + (1-\gamma^y) T_{y+}  + (1+\gamma^y) T_{y-}  + (1-\gamma^z) T_{z+} + (1+\gamma^z) T_{z-} + {} \\ {} + (1-\gamma^\tau)\, \exp\bigg( {i a \Omega_I \frac{\sigma^{12}}{2} } \bigg) T_{\tau+} + (1+\gamma^\tau)\, \exp\bigg(\!{-i a \Omega_I \frac{\sigma^{12}}{2} } \bigg) T_{\tau-}\bigg] - \delta_{x_1,x_2} c_{SW} \kappa \sum_{\mu<\nu} \sigma_{\mu\nu}F_{\mu\nu}\,,
\end{multline*}
where $\kappa = 1/(8+2am)$, $T_{\mu+} = U_\mu(x_1) \delta_{x_1+\mu, x_2}$, $T_{\mu-} = U_\mu^\dagger(x_1) \delta_{x_1-\mu, x_2}$, $F_{\mu\nu} = (\bar{U}_{\mu\nu} - \bar{U}^\dagger_{\mu\nu})/8i $ and
$$\gamma^x = \gamma^1 - {y \Omega_I} \gamma^4,\quad \gamma^y = \gamma^2 +  {x \Omega_I} \gamma^4,\quad \gamma^z = \gamma^3,\quad \gamma^\tau = \gamma^4 .$$
For the clover coefficient $c_{SW}$ we follow Refs.~\cite{Maezawa:2007fc, Ejiri:2009hq} and adopt the mean-field value $c_{SW} =  (1-W^{1\times1})^{-3/4} = (1 - 0.8412/\beta)^{-3/4}$ substituting one-loop result for the plaquette~\cite{Iwasaki:1985we}.

This lattice action (without rotation) has been extensively used by CP-PACS and WHOT-QCD collaborations to study the phase diagram of QCD and its equation of state~\cite{CP-PACS:2000phc, CP-PACS:2000aio, CP-PACS:2001hxw, CP-PACS:2001vqx, Ejiri:2009hq, Maezawa:2007fc}.
The masses of light mesons were calculated for a wide range of simulation parameters in Refs.~\cite{CP-PACS:2000phc, CP-PACS:2000aio, CP-PACS:2001hxw, CP-PACS:2001vqx}, and we reanalyze that data to restore the lines of constant physics (LCP) more frequently in $\beta$ than it was done in Refs.~\cite{CP-PACS:2001hxw, Maezawa:2007fc}. Our interpolation results for the LCPs are consistent with previous studies~\cite{CP-PACS:2001hxw, Maezawa:2007fc} within systematic uncertainties.

Simulations are performed on lattices of the size $N_t\times N_z\times N_s^2$, where $N_s = N_x = N_y$ is the lattice size in directions orthogonal to the rotational axis $z$, which passes through the center of $x,y$-plane.
Due to the metric tensor structure~\eqref{eq:metricE}, we do not encounter a problem with causality in the case of Euclidean rotation, so, in principle, one can consider an unlimited system. The phase diagram of QCD in the presence of fast Euclidean rotation was studied in Refs.~\cite{Chen:2022smf, Chernodub:2022veq}, but the results for an infinite system may not be relevant for the analytical continuation to real angular velocities~\cite{Chernodub:2022qlz}. In present study we focus only on slow rotation and consider the system bounded in $x,y$-directions by the restriction $\Omega_I (N_s-1)a/\sqrt{2} < 1$, and the results then can be analytically continued~\cite{Braguta:2021jgn}.

Since we work with a bounded system, the boundary conditions in corresponding directions $x,y$ are important.
In our previous lattice study of rotating pure Yang-Mills theory several boundary conditions in these directions were implemented (periodic, open, Dirichlet), and it was shown that emerging boundary effects are screened~\cite{Braguta:2021jgn}. Moreover, the critical temperature of the rotating system with different boundary conditions demonstrates qualitatively the same behavior, mainly depending on linear velocity $v_I$ at some point at the boundary, not purely on the angular velocity~\cite{Braguta:2020eis, Braguta:2021jgn, Braguta:2021ucr}. Due to the abovementioned reasons, we conduct this study with periodic boundary conditions (PBC) in $x,y$-directions and fix linear velocity 
    $v_I = \Omega_I (N_s-1) a/2$
in the central point of an outer volume face during simulations. In $z,t$-directions conventional (anti-)periodic boundary conditions are implemented.

\section{Regimes of rotation}
Relativistic rotation enters both parts of the action~(\ref{eq:rot_action_lat_imp_G}),~(\ref{eq:action_F}); thus, several regimes of rotation may be considered to disentangle the effect of rotation on gluons and quarks. For this purpose we introduce separate angular velocities for gluons ($\Omega_G$) and fermions ($\Omega_F$), so the full action is chosen in the form
\begin{equation}
    S = S_G[\Omega_G] + S_F[\Omega_F]\,,
\end{equation}
and the system is investigated in the three following regimes:
\begin{itemize}
    \item only fermionic rotational contribution is probed ($\Omega_G = 0$, \, $\Omega_F=\Omega_I\neq0$);
    \item only gluonic rotational contribution is probed ($\Omega_G =\Omega_I \neq 0$, \, $\Omega_F=0$);
    \item the combined effect is studied ($\Omega_G = \Omega_F = \Omega_I \neq 0$).
\end{itemize}
Of course, in real physical experiments only the last regime can be encountered. Nevertheless, the behavior of the system in different rotational regimes may help to deeper understand rotational phenomena in the theory of strong interactions.

\section{The results}
\begin{figure*}[htb]
\subfigure[]{\label{fig:pl-full-080}
\includegraphics[width=.48\textwidth]{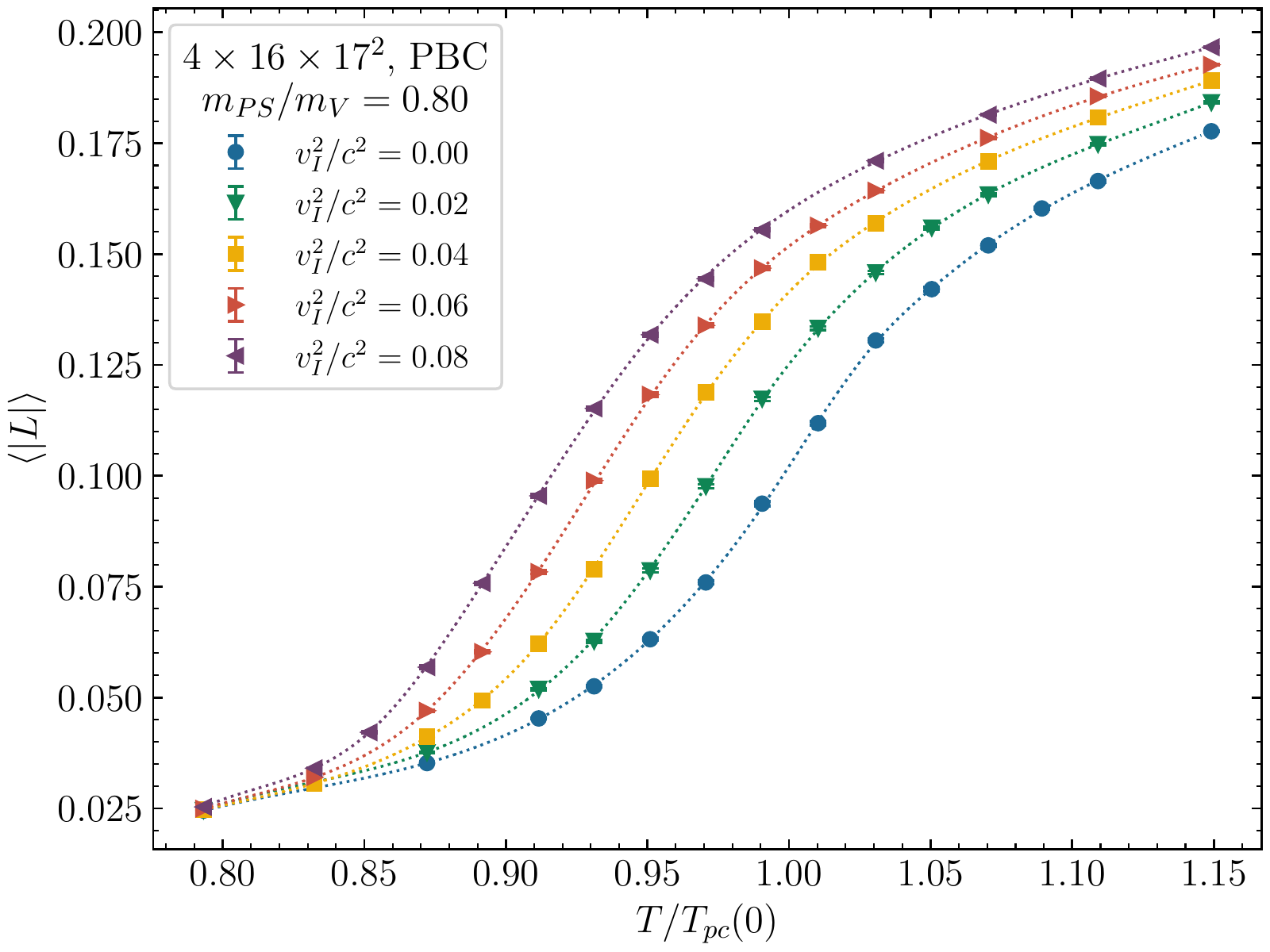}
} \hfill
\subfigure[]{\label{fig:pl-reg-080}
\includegraphics[width=.48\textwidth]{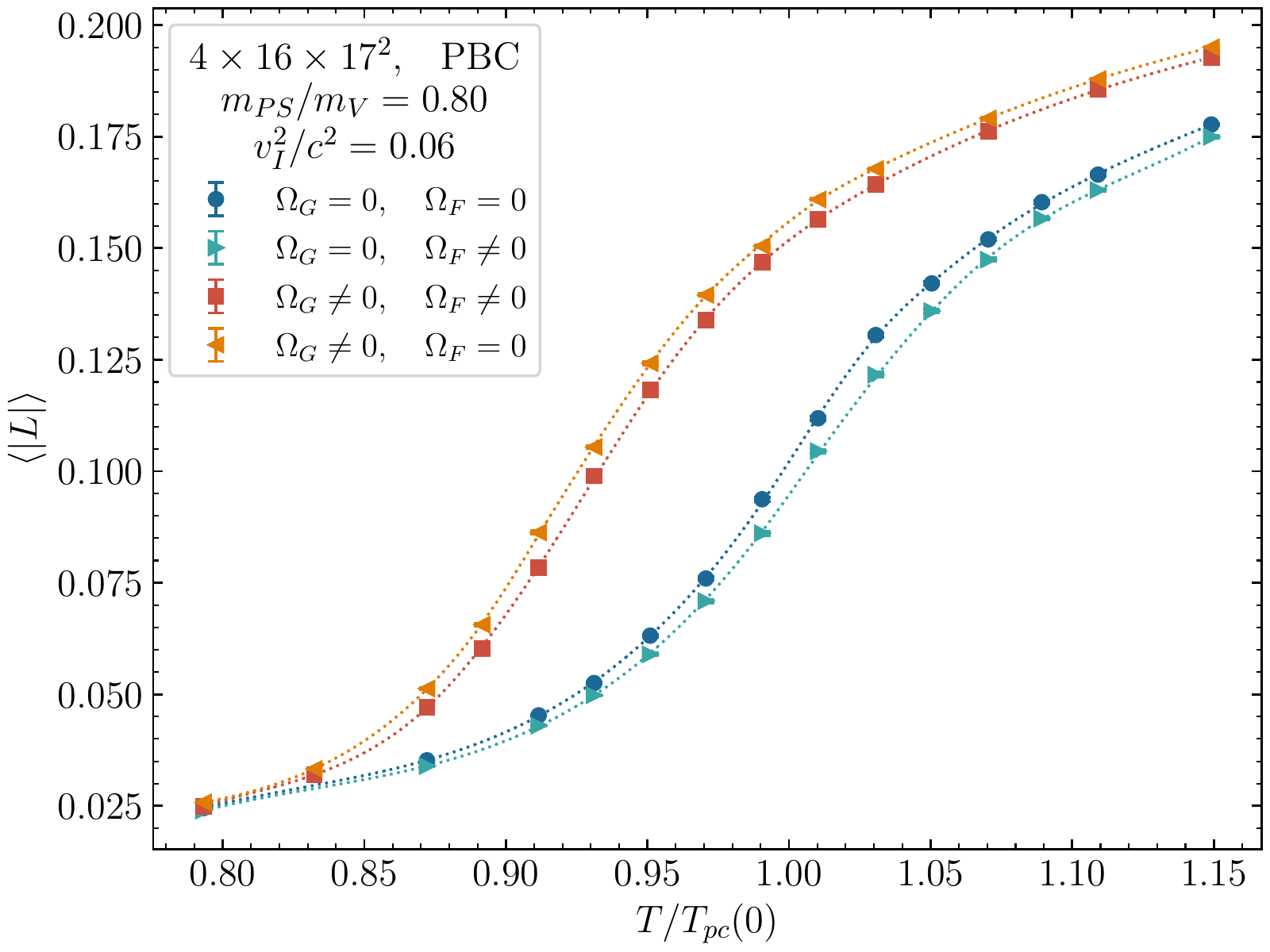}
}
\caption{The averaged Polyakov loop as a function of temperature for different values of imaginary linear velocities at the boundary in the case of full rotation~\subref{fig:pl-full-080}, and for various regimes of rotation at the same value of linear velocity $v_I^2/c^2=0.06$~\subref{fig:pl-reg-080}. The results are obtained on the lattice $4\times 16\times 17^2$ with PBC and $m_{PS}/m_V = 0.80$.}\label{fig:pl-080}
\end{figure*}

Let's consider the regime when both fermions and gluons rotate.
The dependence of the spatially averaged 
Polyakov loop on the normalized temperature $T/T_{pc}(\Omega_I=0)$ for several angular velocities is shown in Fig.~\ref{fig:pl-full-080} for the lattice $4\times 16\times 17^2$ and the ratio of pseudoscalar to vector meson masses $m_{PS}/m_V = 0.80$.
One can see that
the averaged Polyakov loop becomes larger at high temperatures due to rotation,
and the pseudo-critical temperature given by 
the inflection point of the averaged Polyakov loop
decreases with imaginary angular velocity. 

In Fig.~\ref{fig:pl-reg-080} the dependence of the spatially averaged Polyakov loop on the temperature is shown for zero angular velocity as well as for various regimes of rotation with $v_I^2/c^2 = 0.06$. From this plot one can conclude that the shift of the pseudo-critical temperature in the regime when only the gluon action is subjected to rotation ($\Omega_G \neq 0$, $\Omega_F = 0$) is slightly larger than in the regime of full rotation ($\Omega_G = \Omega_F \neq 0$). For the rotational regime when only fermions are subjected to rotation ($\Omega_G = 0$, $\Omega_F \neq 0$) the Polyakov loop becomes smaller at high temperatures, and one could expect that
the pseudo-critical temperature slightly increases due to imaginary rotation.

\begin{figure*}[thb]
\subfigure[]{\label{fig:chi-pl-080}
\includegraphics[width=.48\textwidth]{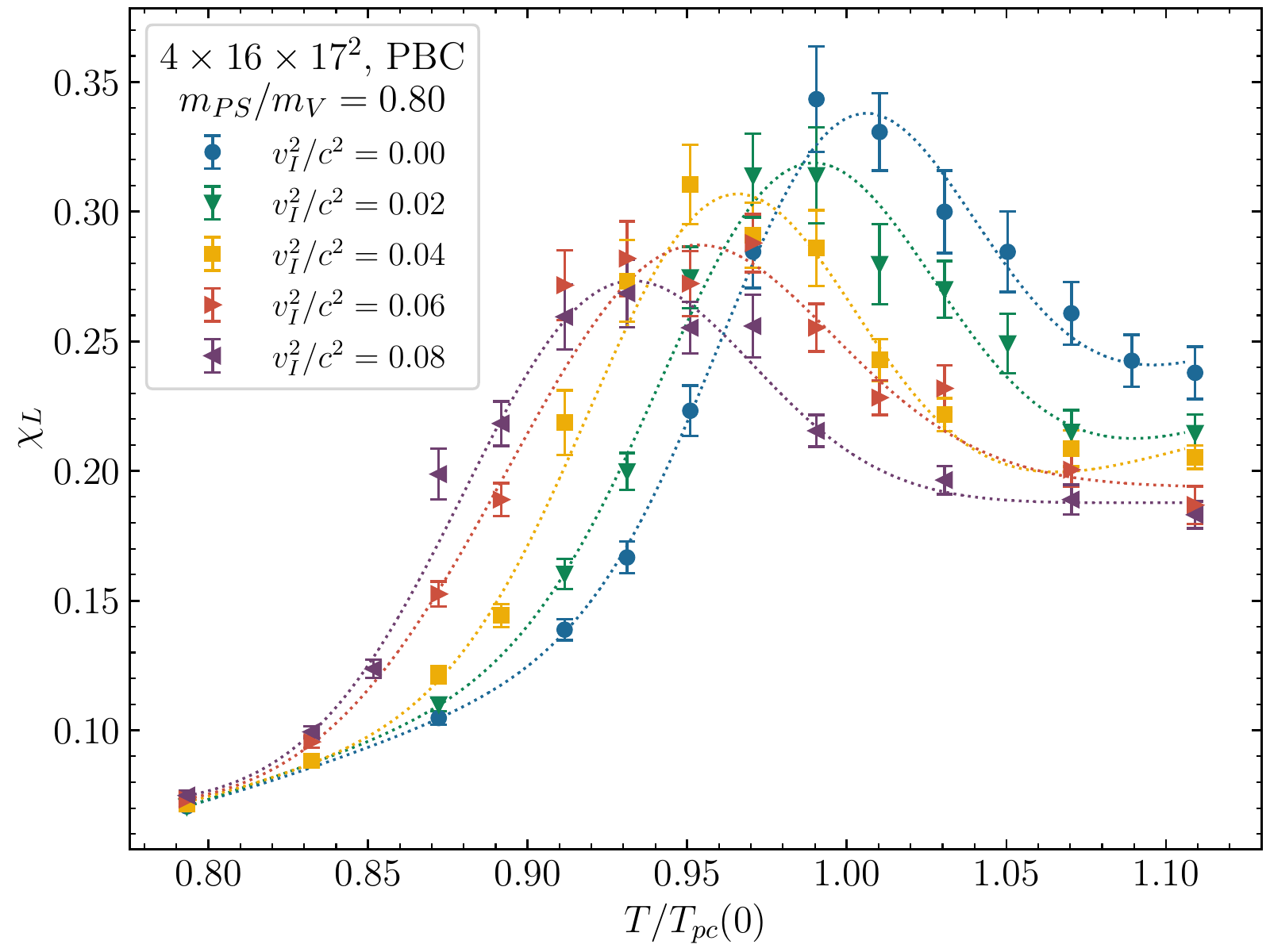}
} \hfill
\subfigure[]{\label{fig:chi-cc-080}
\includegraphics[width=.48\textwidth]{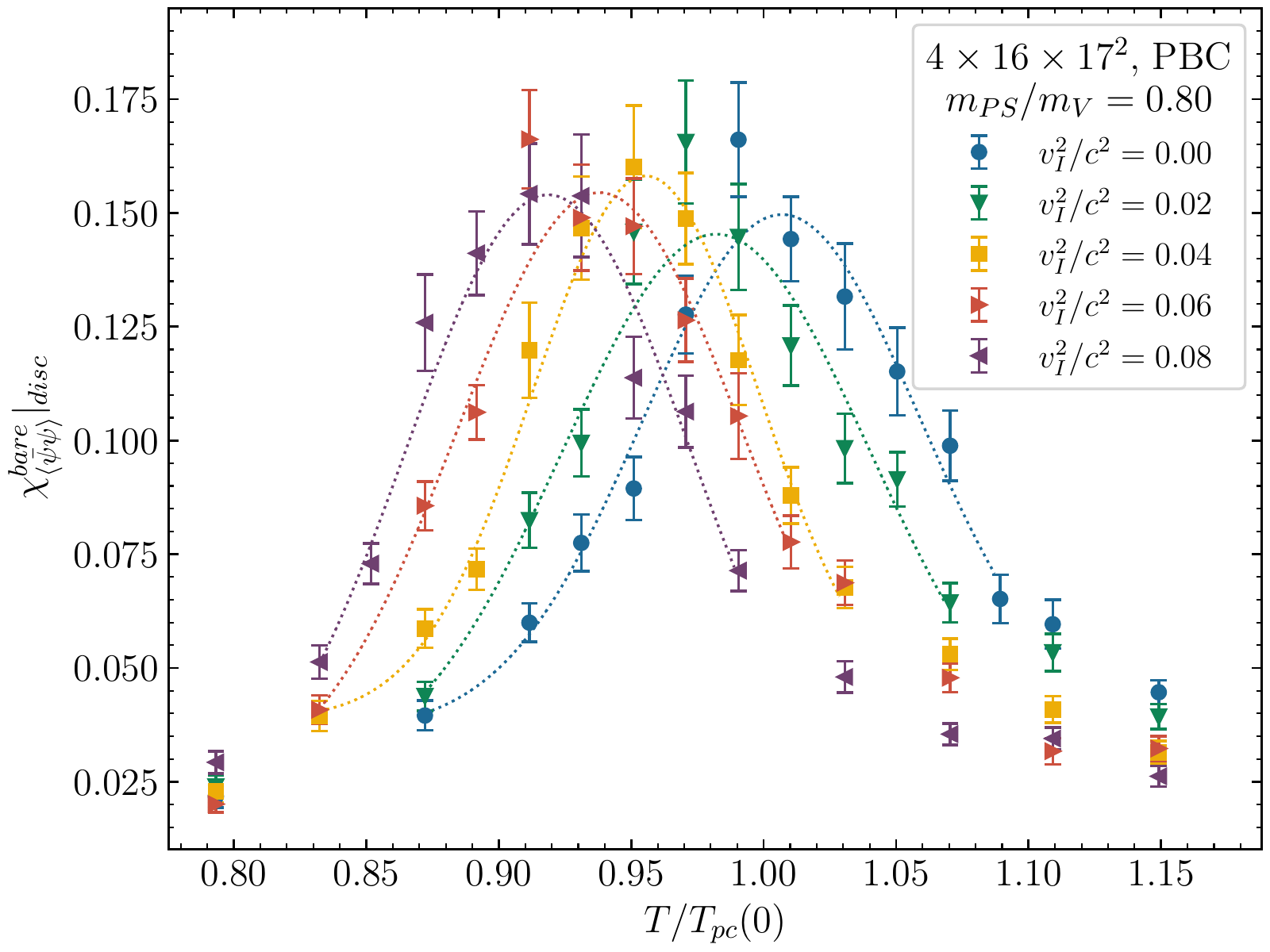}
}
\caption{The susceptibility of the averaged Polyakov loop ~\subref{fig:chi-pl-080} and the (disconnected) chiral susceptibility~\subref{fig:chi-cc-080} as a function of temperature for different values of imaginary linear velocity at the boundary $v_I$ in case of full rotation. The results are obtained on the lattice $4\times 16\times 17^2$ with PBC and $m_{PS}/m_V = 0.80$.}\label{fig:chi-080}
\end{figure*}
\begin{figure*}[htb]
\subfigure[]{\label{fig:tpc-L-reg-080}
\includegraphics[width=.48\textwidth]{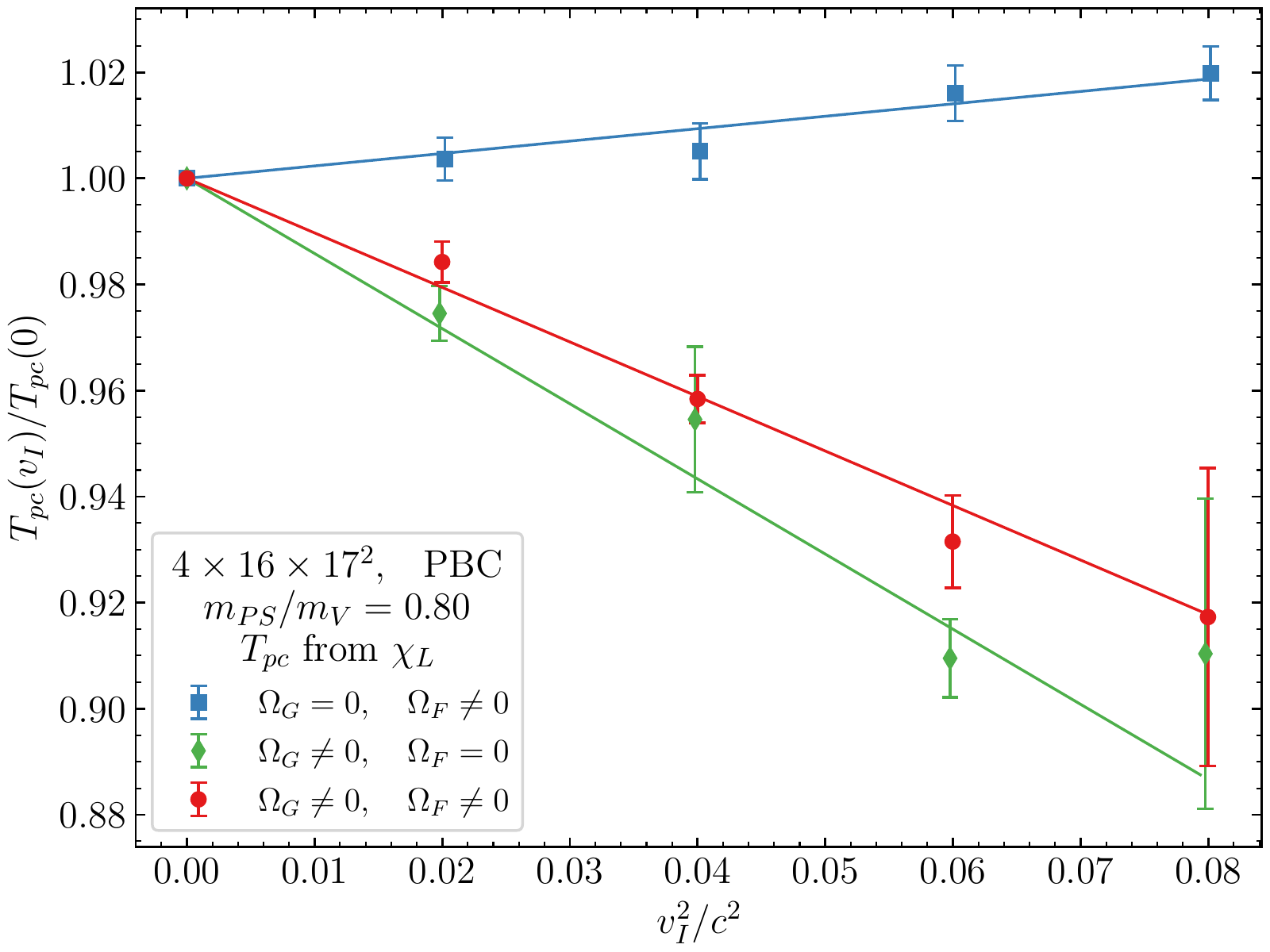}
} \hfill
\subfigure[]{\label{fig:tpc-C-reg-080}
\includegraphics[width=.48\textwidth]{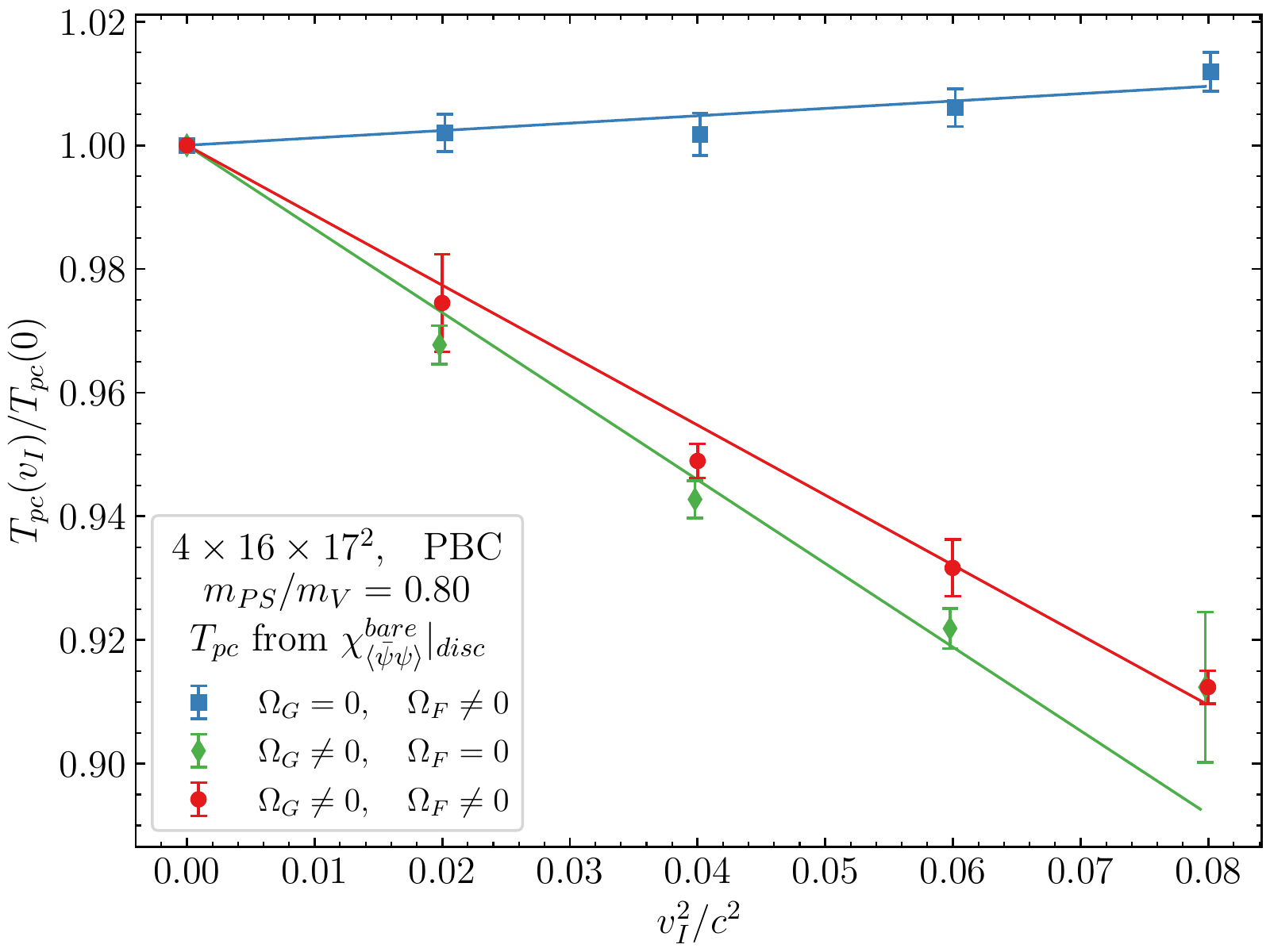}
}
\caption{The pseudo-critical temperature of the confinement-deconfinement~\subref{fig:tpc-L-reg-080} and chiral~\subref{fig:tpc-C-reg-080} crossover as a function of (imaginary) linear velocity squared at the boundary for various regimes of rotation. Lines correspond to the quadratic fits given by Eq.~(\ref{eq:fit_vI}). The results are obtained on the lattice $4\times 16\times 17^2$ with PBC and $m_{PS}/m_V = 0.80$.}\label{fig:tpc-reg-080}
\end{figure*}

In order to make quantitative predictions, we determined the pseudo-critical temperature of the confinement-deconfinement crossover from the position of the peak of the susceptibility of the averaged Polyakov loop. In addition, we measured the pseudo-critical temperature of the chiral crossover from the peak of the disconnected part of the chiral susceptibility. These susceptibilities are shown in Fig.~\ref{fig:chi-080} as functions of temperature for various values of angular velocity in the regime of full rotation.
The resulting values of pseudo-critical temperature vs linear velocity squared $v_I^2/c^2$ for three regimes of rotation and $m_{PS}/m_V = 0.80$ are shown in Figs.~\ref{fig:tpc-L-reg-080},~\ref{fig:tpc-C-reg-080} for both crossovers.

\begin{figure*}[htb]
\subfigure[]{\label{fig:tpc-L}
\includegraphics[width=.48\textwidth]{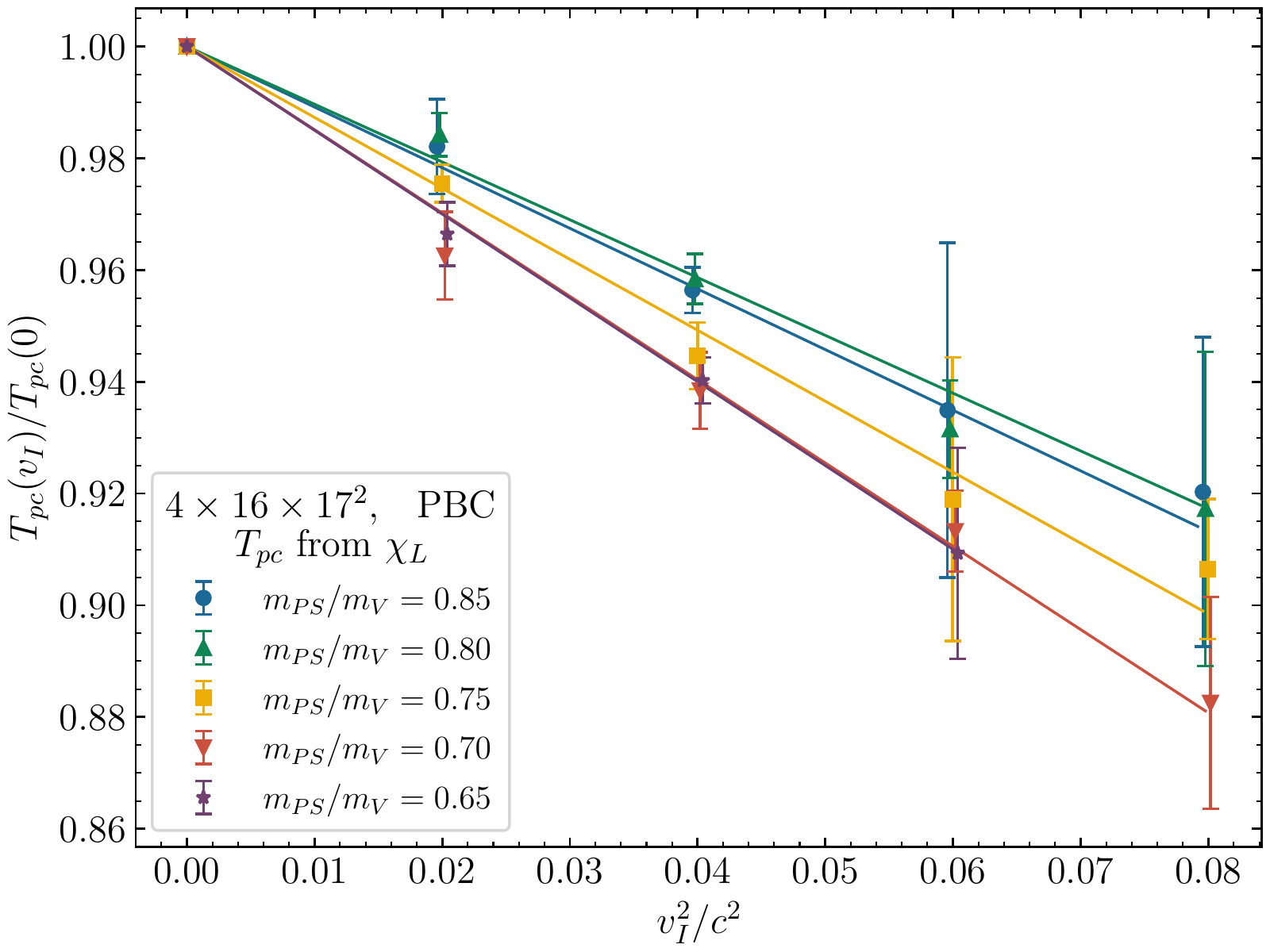}
} \hfill
\subfigure[]{\label{fig:tpc-C}
\includegraphics[width=.48\textwidth]{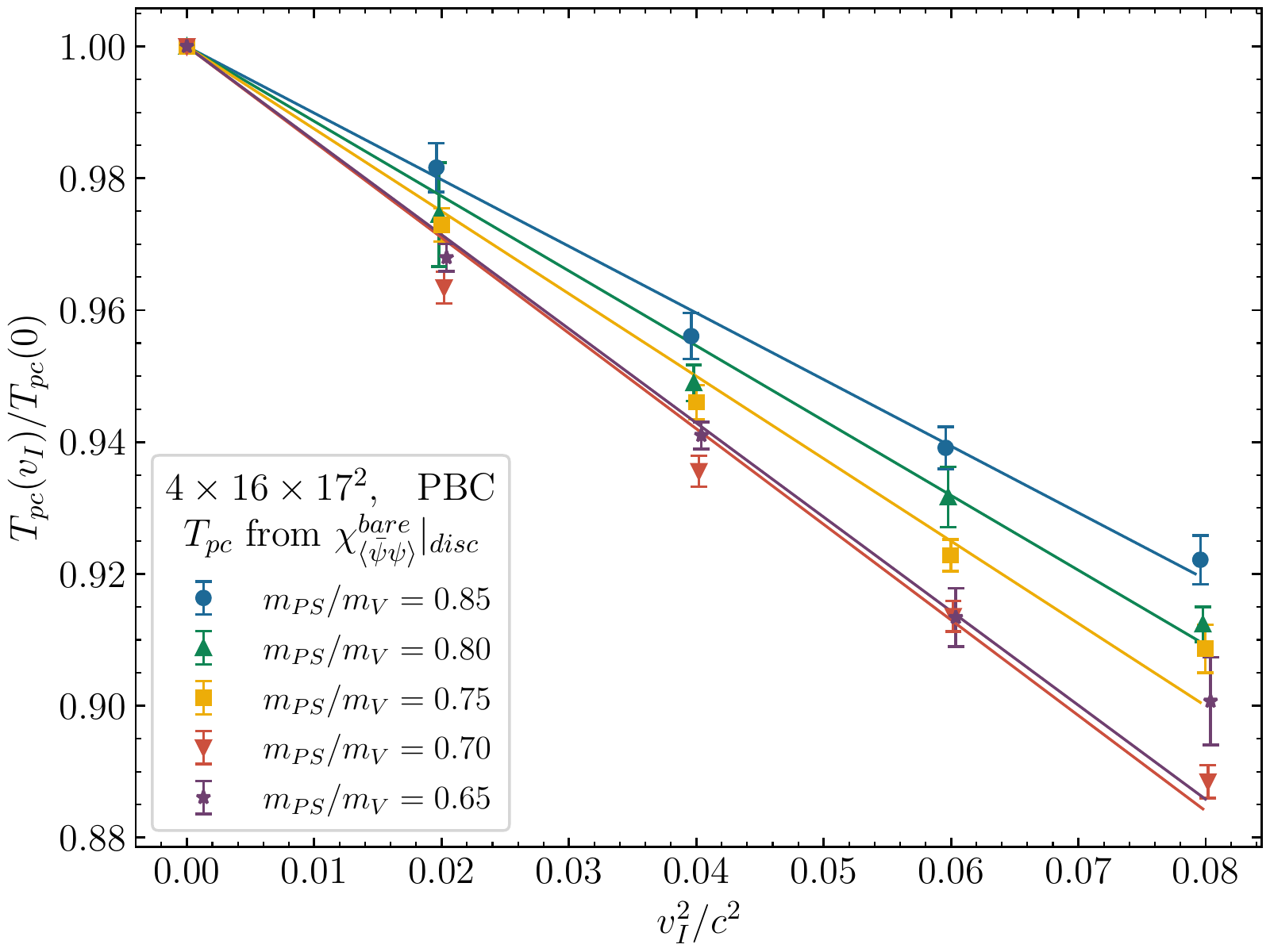}
}
\caption{The pseudo-critical temperature of the confinement-deconfinement~\subref{fig:tpc-L} and the chiral~\subref{fig:tpc-C} crossover as a function of imaginary linear velocity squared at the boundary for various ratios of pseudoscalar to vector meson masses $m_{PS}/m_V$ in the case of full rotation. The results
are from the lattice $4\times 16\times 17^2$ with PBC.}\label{fig:tpc}
\end{figure*}

\vspace{-1.2em}

In Fig.~\ref{fig:tpc-reg-080} all data points with the same regime of rotation are well described by a quadratic function:
\begin{equation}\label{eq:fit_vI}
    \frac{T_{pc}(v_I)}{T_{pc}(0)} = 1 - B_2 v_I^2\, .
\end{equation}
For both deconfinement and chiral crossovers we obtain similar relation between coefficients in Eq.~\eqref{eq:fit_vI} for different regimes: $B_2^{(G)} > B_2 > 0$, $B_2^{(F)} < 0$, where $B_2^{(G/F)}$ is the coefficient for the regime of rotation when only gluon/fermion action includes contribution with non-zero angular velocity, and $B_2$ is the coefficient for the physical regime when all effects are accounted for. One can give an intuitive physical interpretation of these results: separate rotations of fermions and gluons in QCD have opposite effects on the pseudo-critical temperature. But in total, when all parts of the system are subjected to rotation, the pseudo-critical temperatures decrease due to imaginary rotation. It should be noted again that both the chiral crossover and the confinement-deconfinement crossover shift together in the same direction for all rotational regimes.

\begin{figure*}[htb]
\centering
\includegraphics[width=.48\textwidth]{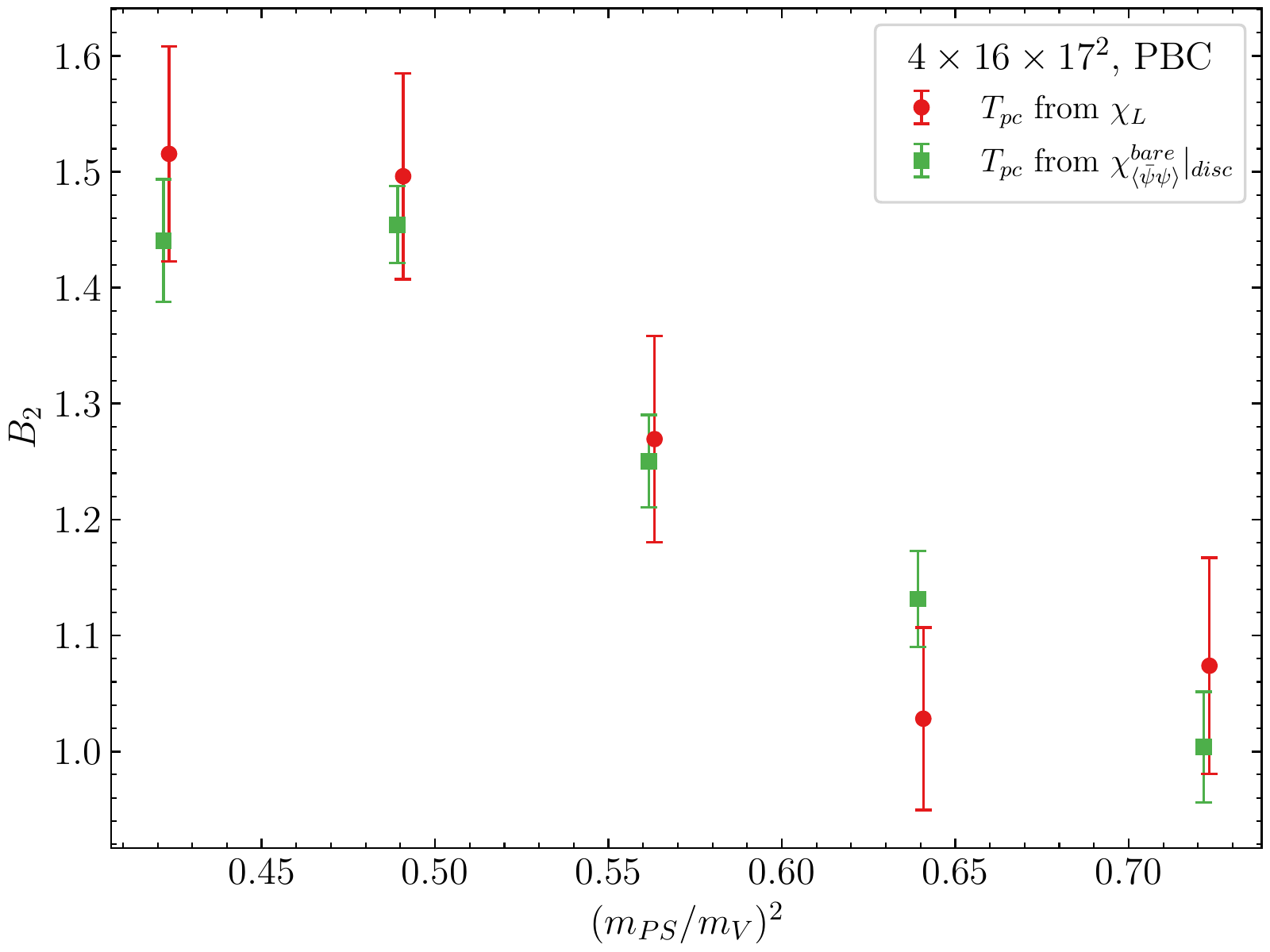}
\caption{The coefficient $B_2$ in Eq.~\eqref{eq:fit_vI} for confinement-deconfinement and chiral crossovers as a function of $(m_{PS}/m_V)^2$. The results are obtained on the lattice $4\times 16\times 17^2$ with PBC.}\label{fig:b2-comp}
\vspace{-.3em}
\end{figure*}

To study how the behavior of pseudo-critical temperatures in rotating QCD depends on the pion mass, we perform simulations for a set of ratios $m_{PS}/m_V$ with full rotation. The results for pseudo-critical temperatures for $m_{PS}/m_V = 0.65, 0.70, 0.75, 0.80, 0.85$ are presented in Fig.~\ref{fig:tpc}, and corresponding coefficients $B_2$ from quadratic fits in Eq.~\eqref{eq:fit_vI} are shown in Fig.~\ref{fig:b2-comp}. Upon analytic continuation to real angular velocity, one can obtain the following equation for both pseudo-critical temperatures: 
\begin{equation}\label{eq:fit_v}
    \frac{T_{pc}(v)}{T_{pc}(0)} = 1 + B_2 v^2\, ,
\end{equation}
where $v = \Omega L_s /2$ and $L_s = (N_s-1)a$ is the size of the system in $x,y$-directions. For all pseudoscalar pion masses we obtained $B_2 > 0$, which means that the pseudo-critical temperatures in QCD increase with real angular velocity, and the coefficient $B_2$ slightly grows when the pion mass becomes smaller.
We are going to extend our study to smaller pion masses and finer lattices in forthcoming works.

\section{Conclusions}
\vspace{-.2em}

We presented our first results for the phase diagram of rotating QCD with $N_f=2$ clover-improved Wilson fermions. Lattice simulations were performed in rotating coordinates, where rotation was introduced via an external gravitational field.
In order to overcome the sign problem, the system was simulated at imaginary angular velocity. The pseudo-critical temperature of the confinement-deconfinement crossover as well as the temperature of the chiral crossover were shown to decrease with imaginary rotation. After the analytic continuation to real angular velocity we found the pseudo-critical temperatures to increase quadratically with angular velocity (see Eq.~\eqref{eq:fit_v}), with the coefficients of proportionality being consistent within statistical uncertainties (see Fig.~\ref{fig:b2-comp}). For all considered values of the pion mass ($m_{PS}/m_V = 0.65, \dots, 0.85$) the pseudo-critical temperatures in rotating QCD were found to increase with rotation, likewise in SU(3) gluodynamics~\cite{Braguta:2020eis, Braguta:2021jgn, Braguta:2021ucr}.

Moreover, we considered several regimes of rotation. It was shown that the separate rotations of fermions and gluons have opposite influences on the critical temperatures: rotating gluons tend to increase it, whereas rotating quarks lead to its decrease.
These results are in agreement with our first results for standard Wilson fermions reported in Ref.~\cite{Braguta:2021ucr} and may be related to the mechanism proposed in Ref.~\cite{Jiang:2021izj}. It should be noted that the standard NJL model (with rotation-independent coupling)
predicts a decline in the critical temperature~\cite{Jiang:2016wvv, Ebihara:2016fwa, Wang:2018sur, Chernodub:2016kxh}.
Our results for different regimes of rotation demonstrate the behavior of rotating QCD
to be more complicated than kinematic predictions from general relativity, i.e. the Ehrenfest-Tolman effect (see discussions in Refs.~\cite{Braguta:2020eis, Braguta:2021jgn, Braguta:2021ucr, Chernodub:2020qah, Chernodub:2022veq, Chernodub:2022qlz}).
This suggests an importance of the contribution from rotating gluons to fully understand the properties of rotating QCD medium.
\vspace{-.3em}

\acknowledgments
\vspace{-.2em}

This work has been carried out using computing resources of the Federal collective usage center Complex for Simulation and Data Processing for Mega-science Facilities at NRC ``Kurchatov Institute'', \url{http://ckp.nrcki.ru/}  and the Supercomputer  ``Govorun'' of Joint Institute for Nuclear Research.

\vspace{-.3em}
\bibliographystyle{JHEP}
\bibliography{biblio.bib}

\end{document}